\documentclass[12pt,fleqn]{article}

\usepackage{epsfig}
\usepackage{graphics}
\usepackage{latexsym}
\usepackage{amssymb}
\usepackage{amsmath}
\def\one{{\sf 1}\mkern-5.0mu{\rm I}}
\newcommand{\ben}{\begin{displaymath}}
\newcommand{\een}{\end{displaymath}}

\title{Asymptotic Factorisation of the Ground-State for SU(N)-invariant
Supersymmetric Matrix-Models}
\author{  D. Hasler${}^{(a)}$,  J.
Hoppe${}^{(b)}$ \\ 
\vspace*{-0.05truein} \\
\normalsize\it ${}^{(a)}$ Theoretische Physik,
ETH-H\"onggerberg, CH--8093 Z\"urich\\
\normalsize\it   ${}^{(b)}$ Department of
Mathematics, KTH, S-10044 Stockholm}

\begin{document}

\maketitle
\vspace{0.4cm}
\begin{abstract}
We give a simple - straightforward and rigorous - derivation that when the
eigenvalues of one of the $d=9 \ (5,3,2)$ matrices in the $SU(N)$ invariant
supersymmetric matrix model become large (and well separated from each other)
the ground-state wavefunction (resp. asymptotic zero-energy solution of the
corresponding differential equation) factorizes, for all $N>1$, into a product
of supersymmetric harmonic oscillator wavefunctions (involving the
`off-diagonal' degrees of freedom) and a wavefunction $\psi$ that is
annihilated by the free supercharge formed out of all `diagonal' (Cartan
sub-algebra) degrees of freedom.
\end{abstract}

During the past few years, zero-energy states in supersymmetric matrix-models
have been widely investigated [1-13]. In this paper, we will derive the
asymptotic form of the ground-state wavefunction, for arbitrary $N>1$. While
the result we obtain may not be surprising (A. Smilga has
previously stated the emergence of effectively free asymptotic
supercharges, referring to work of E. Witten, and himself (cf.[14][15]); and the
free Laplacian in [10] should come from an effective, hence also free
supercharge) it is perhaps worth giving an explicit proof of how asymptotic
solutions of 
\ben
Q_{\beta} \Psi = 0 \ ,
\een 
\begin{eqnarray} \label{Q}
Q_{\beta}  =  \left( -i\frac{\partial}{\partial q_{tA}} \gamma^{t}_{\beta \alpha}
+ \frac{1}{2} f_{ABC} q_{sB}q_{tC}\gamma^{st}_{\beta \alpha} \right)
\Theta_{\alpha A}  \\
\nonumber
 A,B,C = 1, ...,N^2 -1,  \\ \nonumber
 s,t = 1, ...,d \, (d=2,3,5 \ \textrm{or}\  9), \\ \nonumber
 \alpha,
\beta = 1, ...,2(d-1)
\end{eqnarray}
look like; $f_{ABC}$ are real, totally antisymmetric structure constants of
$SU(N)$, the hermitian fermionic operators $\Theta_{\alpha A}$ satisfy
canonical anticommutation relations, $\{  \Theta_{\alpha A}, \Theta_{\beta
B} \} 
= \delta_{\alpha \beta} \delta_{A B}$, and the real symmetric
$\gamma$-matrices satisfy
\begin{equation} \label{gamma}
\gamma^{s}_{\alpha \beta}\gamma^{st}_{\alpha' \beta'} + 
\gamma^{s}_{\alpha' \beta}\gamma^{st}_{\alpha \beta'} + (\beta \leftrightarrow
\beta') 
= 
2 \left( \delta_{\alpha \alpha'} \gamma^{t}_{\beta \beta'} - \delta_{\beta
\beta'} \gamma^{t}_{\alpha \alpha'} \right) 
\end{equation}
in addition to $\gamma^s \gamma^t + \gamma^t \gamma^s = 2 \delta^{ts} \cdot
\one$. On SU(N) invariant states, 
\begin{equation} \label{sun}
J_{A} \Psi := -i f_{ABC} \left( q_{sB} \frac{\partial}{\partial q_{sC}} +
\frac{1}{2} \theta_{\alpha B} \theta_{\alpha C} \right) \Psi = 0 \ ,
\end{equation}
letting $A=k=1, ...,N-1$ label a Cartan subalgebra (CSA) (and $A=a=N, ..., N^2
-1$ the non-CSA degrees of freedom), fixing the gauge by choosing
\begin{equation} \label{gaugecond}
q_{da}=0 \ \ (a=N, ..., N^2-1) 
\end{equation}
as well as assuming
\begin{equation} \label{5}
f_{ab} := \sum_{k=1}^{N} f_{abk} q_{dk}
\end{equation}
to be invertible ($f:= \det (f_{ab}) \neq 0$),
\ben
\tilde{Q}_{\beta} := \sqrt{f} Q_{\beta} \frac{1}{\sqrt{f}}
\een
may be written as $(\mu =1,2,...,d-1)$
\begin{eqnarray}  \label{6}
\tilde{Q}_{\beta} = 
&  & \left( -i \nabla_{\mu a} \gamma^{\mu}_{\beta \alpha} +
f_{ca} q_{\mu c} \gamma_{\beta \alpha}^{d \mu} \right) \Theta_{\alpha a} \\
\nonumber 
& + & f_{akc}q_{\mu k} q_{\nu c} \gamma_{\beta \alpha}^{\mu \nu}
\Theta_{\alpha a}  \\ \nonumber
& - & i \nabla_{\mu k} \gamma^{\mu}_{\beta \alpha} \Theta_{\alpha k} - i
\nabla_{d k} 
\gamma_{\beta \alpha}^{d} \Theta_{\alpha k} \\ \nonumber
& - & \frac{i}{2} \gamma^{d}_{\beta \alpha} \Theta_{\alpha k} f^{-1}_{ab}
f_{abk}
+ i \gamma_{\beta \alpha}^{d} \Theta_{\alpha a} f^{-1}_{ab} \left( -
\frac{1}{2} 
f_{bCD} \Theta_{\alpha' C} \Theta_{\alpha' D}  - i {L'}_b \right) \\
\nonumber 
& + & \frac{1}{2} f_{abc} q_{\mu b} q_{\nu c} \gamma^{\mu \nu}_{\beta \alpha}
\Theta_{\alpha a} + \frac{1}{2} f_{kbc}q_{\mu b}q_{\nu c} \gamma^{\mu
\nu}_{\beta \alpha} \Theta_{\alpha k} \ ,
\end{eqnarray}
where the first term in line 4 comes from
\ben
\frac{\partial}{\partial q_{dk}} \sqrt{f} = \frac{1}{2} \sqrt{f} f^{-1}_{ab} 
\frac{\partial}{\partial q_{dk}} f_{ba} \ ,
\een
and the second one (with $i{L'}_b = \sum_{\mu = 1}^{d-1} f_{bCE} q_{\mu C}
\nabla_{\mu E}$) results upon using (\ref{sun}) to replace the terms involving
\footnote{Sorry for the notation: d is always 2,3,5 or 9, and never
corresponds to an $SU(N)$ index.} $\nabla_{da}$ when acting on gauge-fixed
functions (cp. [16])
\ben
\tilde{\Psi} = ({\mathrm{const}}) \cdot \sqrt{f} \cdot \Psi ( \ldots  q_{da} \equiv 0
\ldots ) \ ;
\een
the factor $\sqrt{f}$ arises as follows: with $X_{s} := q_{sA}T_{A}$,
$\left\{ T_{A} \right\}_{A=1}^{N^2-1}$ being traceless hermitian $N \times N$
matrices (diagonal for $A=1,\ldots,N-1$), satisfying  $\left[ T_A , T_B
\right] = i f_{ABC} T_C$, (\ref{sun}) says that for unitary $U$
\ben
\Psi(X_1, \ldots , X_d ) = \hat{U} \Psi ( U^{-1} X_1 U , \ldots , U^{-1} X_{d}
U ) \ ,
\een
with $\hat{U}$ representing the action of $U$ in the fermionic Fock space; as
it is well known that when changing coordinates from a hermitian matrix 
($X_d,$ resp. the $q_{dA}$) to its eigenvalues $\lambda_1 >  \lambda_2 \ldots
>  \lambda_N$ and the parameters of the unitary matrix $U$ diagonalizing it,
$X_d = U \Lambda U^{-1}$, the Jacobian of the transformation is $\prod_{i < j}
(\lambda_i - \lambda_j )^2$, while $\prod_{i=1}^{N} d\lambda_i \delta(\sum
\lambda_i ) \sim \prod_{i=1}^{N-1} q_{dk}$, we can consider $X_d$ to be equal
to $\Lambda$ (this is the content of (\ref{gaugecond})), while absorbing the
Jacobian (and the constant $\int{dU}$ ), 
\ben
\int{\prod dq_{\mu A}} \int{ \prod dq_{dA}} = \int{ \prod dq_{\mu A}} \int{dU}
\int{dq_{dk} \prod_{i<j} (\lambda_i - \lambda_j)^2} \ , 
\een
that enter the calculation of scalar products of gauge-invariant states into
the gauge-fixed wave functions. 
In order to see that $\prod_{i<j}(\lambda_i - 
\lambda_j )$  is equal to $\sqrt{f}$, one notes that $\left[ \Lambda , E_{jk}
\right] = (\lambda_j - \lambda_k ) E_{jk}$ for $(E_{jk})_{mn} = \delta_{jm}
\delta _{kn}$, while writing $E_{jl} = e_{jl}^{a} T_a$ and putting $\Lambda =
q_{dk} T_k$ gives 
\begin{eqnarray} \label{7}
(\lambda_j - \lambda_k ) e_{jk}^{a} T_a & = & \left[ \Lambda, E_{jk} \right] =
q_{dk}e_{jk}^{a} \left[ T_k, T_a \right] \\ \nonumber
& = & i q_{dk} f_{kab} e_{jk}^a T_b = \left( - i f_{ab} e_{jk}^b \right) T_a \
, 
\end{eqnarray}
hence $f_{ab}e_{jk}^b = i ( \lambda_j - \lambda_k ) e_{jk}^a $; so the product
of eigenvalues of the matrix $(f_{ab})$ is
\ben
\prod_{j \neq k} i (\lambda_j - \lambda_k ) = {\left( \prod_{j<k} ( \lambda_j -
\lambda_k ) \right)}^2 \ ;
\een
note that the eigen\emph{vectors} of $(f_{ab})$ are independent of the
parameters $q_{dk}$ (we will make use of that later).  
Returning to equation (\ref{6}) let us denote the first line by $Q_{\beta}^0$,
the third line by 
$\hat{Q}_{\beta}$, and the rest by $Q'_{\beta}$.
Using (\ref{gamma}) when 
calculating the cross-terms it is straightforward to see that 
\begin{eqnarray} \label{8}
\left\{ Q_{\beta}^0 , Q_{\beta'}^0 \right\} &  = & \delta_{\beta \beta'}
H^{(0)} + 
2 
(\gamma^d)_{\beta \beta'} q_{dk} J_{k} \ ,  \\ \nonumber
H^{(0)} & = & -\nabla_{\mu a} \nabla_{\mu a} + f_{ca} f_{cb} q_{\mu a} q_{\mu
b} + 
i f_{ca} \gamma^d_{\alpha \alpha'} \Theta_{\alpha c} \Theta_{\alpha' a} \ .
\end{eqnarray}
Both the bosonic and the fermionic part of $H^{(0)}$ can easily be
diagonalized by using $(if_{ac}) e_{jk}^c = - (\lambda_j - \lambda_k )
e_{jk}^c $, resp. writing $(if_{ab}) = UDU^{-1}$, with the columns of the
unitary matrix $U$ being the eigenvectors $\vec{e}_{jk}$; so, defining
\begin{eqnarray} \label{d} 
z_{\mu , jk} & := & {(e_{jk}^a )}^{\ast}q_{\mu a} \ \left( = e_{kj}^a q_{\mu a}
\right) \\ \nonumber
\Theta_{\alpha ,jk}' & := & {(e_{jk}^a)}^{\ast} \Theta_{\alpha a}  \ \left( =
e_{kj}^{a} \Theta_{\alpha a} \right) \ ,
\end{eqnarray}
$H^{(0)}$ becomes $2 \cdot \sum_{j<k} H_{jk}^{(0)}$,
\begin{eqnarray} \label{10}
H_{j<k}^{(0)}  = & - & \frac{\partial}{\partial z^{\ast}_{\mu,jk}} 
                  \frac{\partial}{\partial z_{\mu,jk}} + (\lambda_j -
\lambda_k )^2  z^{\ast}_{\mu,jk}  z_{\mu,jk}   \\ \nonumber
& - & \gamma^d_{\alpha \alpha'} (\lambda_j - \lambda_k )
{\Theta'}^{\dagger}_{\alpha, 
jk} {\Theta'}_{\alpha' , jk} \ ,
\end{eqnarray}
whose ground state (now choosing $\gamma^{d}$ as 
$\left( \begin{array}{cc} \one & 0 \\ 0 \ & - \one \end{array}
\right)$)
\begin{equation} \label{11}
\psi_0 = \prod_{j<k} { {\left( \frac{2}{\pi} \right)}^{\frac{d-1}{2}}(
\lambda_j - \lambda_k )}^{\frac{d-1}{2}} \cdot  e^{ - 
( \lambda_j - \lambda_k ) z^{\ast}_{\mu ,jk} z_{\mu  
, jk}}  \cdot \prod_{\alpha = 1}^{d-1} {\Theta'}^{\dagger}_{\alpha ,jk} \mid 0
\rangle   \ ,
\end{equation}
where the vacuum is defined by
\ben
 {\Theta'}_{\alpha ,jk} \mid 0
\rangle  = 0  \ , \ \  \forall \alpha , \ \forall (j,k) \ \mathrm{with} \ j< k , 
\een
has energy zero (and vanishes when any of the $\lambda_j$ become equal). Now
assume that the $q_{dk}$ and the eigenvalues of $(f_{ab}) = (f_{abk}q_{dk})$
all become large, of order $r$ $(r \to \infty)$, with the $z_{\mu ,jk}$ and
the $q_{\mu a}$ small, of order $\frac{1}{\sqrt{r}}$ (cp. (\ref{11})), while
the $q_{\mu k}$ are considered to be of order $1$. The leading part in
(\ref{6}) is then $Q_{\beta}^{(0)}$, being of order $\sqrt{r}$. Denoting the
scalar-product with respect to the off-Cartan degrees of freedom, $q_{\mu a}$
and $\Theta_{\alpha a}$, by round brackets, we will now show that
$\left( \psi_{0}, Q'_{\beta} \psi_{0} \right) = 0$, so that $\psi$
(depending on the Cartan-degrees of freedom) in $\tilde{\Psi} \sim \psi_0 \cdot \psi$
$(r \to \infty)$ has to satisfy $\hat{Q}_{\beta} \psi = 0$. All terms linear (or
cubic) in the off-Cartan fermions clearly give zero, when calculating the
expectation value. The last term in (\ref{6}) also does not contribute (upon
$\int{\prod dq_{\mu b}}$), as necessarily $\mu \neq \nu$ (and $b \neq c$), so
it is odd in $q_{\mu b}$, resp. $q_{rc}$. The term which is cubic in the
fermions, contributing via 
\ben
+ i \gamma^{d}_{\beta \alpha} \Theta_{\alpha' k} f^{-1}_{ab} f_{cbk} \left(
\psi_0 , \Theta_{\alpha a} \Theta_{\alpha' c} \psi_0 \right)
\een
cancels the contribution from the preceding term as
\ben
\begin{array}{ccc}
\left(
\psi_0 , \Theta_{\alpha a} \Theta_{\alpha' c} \psi_0 \right) & = & \frac{1}{2}
\delta_{\alpha \alpha'} \delta_{ac} \left( \psi_0 , \psi_0 \right)  \\
||     &  &  || \\
e^{a}_{jk} e^{c}_{ln} \left( \psi_0 , {\Theta'}_{\alpha , jk}
{\Theta'}_{\alpha' 
,  
ln} \psi_{0} \right) & = & \frac{1}{2} e^{a}_{jk} e^{c}_{kj} \delta_{\alpha
\alpha} \left( \psi_0 , \psi_0 \right) 
\end{array}
\een
due to $UU^{\dagger} = \one$.
\newline
Let us discuss the contribution of excited states, i.e. states orthogonal to
$\psi_0$.  We define the projection operator $P_0$, which projects onto the
state $\psi_0$ as
\ben
P_0 \tilde{\Psi} = \psi_0 ( \psi_0 , \tilde{\Psi} ) \ ,
\een 
and $\overline{P}_0 = 1 - P_0$. Let $\Psi$ be a zero mode of
$Q_{\beta}$, i.e.
$Q_{\beta} \Psi = 0$, or equivalently
\ben 
\tilde{Q}_{\beta}
\tilde{\Psi} = 0.
\een
Decomposing this equation with respect to $P_0$ and $\overline{P}_0$, we 
obtain 
\ben 
\left( \begin{array}{cc}  
P_0 \tilde{Q}_{\beta} P_0 & P_0 \tilde{Q}_{\beta} \overline{P}_0 \\ 
\overline{P}_0 \tilde{Q}_{\beta} P_0 & \overline{P}_0 \tilde{Q}_{\beta} 
\overline{P}_0   
\end{array} \right)  
\left( \begin{array}{c} 
P_0 \tilde{\Psi} \\ 
\overline{P}_0 \tilde{\Psi} \end{array}
\right) = 0 \ . 
\een 
The scaling behavior for $r \to \infty$ of the operators $\tilde{Q}_{\beta}$
is given by
\ben 
\left( \begin{array}{cc}  
P_0 \tilde{Q}_{\beta} P_0 & P_0 \tilde{Q}_{\beta} \overline{P}_0 \\ 
\overline{P}_0 \tilde{Q}_{\beta} P_0 & \overline{P}_0 \tilde{Q}_{\beta} 
\overline{P}_0   
\end{array} \right) =  
\left( \begin{array}{cc}  
O(r^{0}) & O(r^{-1/2}) \\ 
 O(r^{-1/2}) &  O(r^{1/2}) 
\end{array} \right)   \ .
\een  
Note that the term of order $r^{1/2}$ in $\overline{P}_0 \tilde{Q}_{\beta} 
\overline{P}_0 $ is invertible on the range of $\overline{P}_0$. We obtain 
\ben 
\begin{array}{cc} 
O(r^0) P_0 \tilde{\Psi} + O(r^{-1/2})\overline{P}_0 \tilde{\Psi} & = 0 \\ 
O(r^{-1/2}) P_0 \tilde{\Psi} + O(r^{1/2})\overline{P}_0 \tilde{\Psi} & = 0  
\end{array} 
\een 
The second equation shows that the term $\overline{P}_0 \Psi$ is 
suppressed with respect to $P_0 \Psi$ at least by a factor $r^{-1}$. It 
follows that 
\ben 
P_0 \tilde{Q}_{\beta} P_0 \tilde{\Psi} = 0 + O(r^{-3/2}) \ . 
\een 
We identify the range of $P_0$ with the space of effective wavefunctions 
\ben 
\psi = \psi(q_{\mu k} , \Theta_{\alpha k} ) 
\een 
via 
\ben 
P_0 \tilde{\Psi} \mapsto \psi = ( \psi_0 , \tilde{\Psi} )  \ .
\een 
Hence $P_0 \tilde{\Psi} = \psi_0 \cdot \psi$.  Let us summarize the above
result. To a 
solution  
$\Psi$ of $Q_{\beta} \Psi = 0$, there corresponds a wavefunction $\psi$ given
by $P_0 
\tilde{\Psi} = \psi_0 \cdot \psi$ such that 
\ben 
\hat{Q}_{\beta} \psi = 0 + O(r^{-3/2}) \ \mathrm{for} \ \ r \to \infty \ ;
\een 
$\psi$ must be square integrable at infinity.
\\  \\
\medskip  
\noindent {\bf Acknowledgments.\/} This work was done while one of us (D.H.)
was visiting KTH. We thank ETH and KTH for financial support,
resp. hospitality, - and G.M. Graf for discussions.


\end{document}